\documentclass[letterpaper
,aps,pra
,twocolumn
,showkeys
]{revtex4-2}

\usepackage{graphicx}
\usepackage{dcolumn}
\usepackage{bm}
\usepackage{latexsym}
\usepackage{ifsym}
\usepackage{stix}
\usepackage[utf8]{inputenc}
\usepackage{amsmath}
\usepackage{longtable}
\usepackage{ltxtable}
\usepackage{booktabs}

\begin{document}

\bibliographystyle{apsrev}


\title{Calculation of the K $4s, 7s,$ and $4p$ photoabsorption near threshold}

\author{Constantine E. Theodosiou}
\affiliation{Department of Physics and Astronomy, Manhattan College, Riverdale, New York 10471}
\email{E-mail: constant.theodosiou@manhattan.edu}

\date{\today}            

\begin{abstract}
We revisit the photoabsorption from the ground state of K, ending below and above the ionization threshold, with special emphasis on the shape of the photoionization cross section around the Cooper minimum.  The present treatment, including core polarization effects and dynamic core polarizability matches very well the most accurate measurements resolving an older question about the width of the Cooper minimum. 
The  photoabsorption from the excited $7s$ and $4p$ states is also calculated,  along with $4p$$\rightarrow $$nl$ optical oscillator strengths, and compared with recent   experimental and theoretical data.
\end{abstract}

\pacs{...}

\maketitle


\section{Introduction} 
The ground-state photoabsorption cross sections of alkali metal atoms are known to have distinct minima \cite{Hudson1965, Marr1968, Ditchburn1953}, called Cooper minima
\cite{Cooper1962, Fano1968}, just above the ionization threshold for Na, K, Rb, and Cs, and below threshold in Li. This is an interesting behavior since the photoabsorption of H, which they closely resemble in terms of their spectra does not exhibit such minima. Their origin lies in the fact that the alkali metal atoms have a finite core that perturbs the otherwise hydrogenic wavefunctions of the outer electrons enough to cause enough ``out-of-phase" shift and change in the initial-final state wavefunction overlap that at the appropriate energy the photoabsorption transition matrix element goes through a sign change.  The resulting transition matrix element zeroes do not result to actual cross section cancellations because the effect of spin-orbit interaction shifts differently the $\epsilon p_{1/2}$ and $\epsilon p_{3/2}$ so that the partial waves have actual zeroes but the sum of the partial cross sections have non-zero, albeit of small magnitude, minima in the observed cross sections.
The width and depth of these minima change with the atomic number as the spin-orbit interaction also changes correspondingly \cite{Seaton1951}. Potassium has the additional interesting feature that does not have a strong enough core to support a $d$ orbital in its ground state which creates a type of instability in the formation of excited states.

There is an extensive number of theoretical calculations for the photoionization of alkali metal vapors above the first ionization threshold.  In the case of potassium they include Hartree-Slater \cite{Fano1968}, Hartree-Fock \cite{Chang1975}, many-body perturbation \cite{Chang1975}, multi-configuration Hartree-Fock \cite{Saha1989}, relativistic random-phase approximation \cite{Fink1986}, semi-empirical \cite{Weisheit1972, Marinescu1994}, and more recently configuration interaction Pauli-Fock approach with core 
polarization potentials \cite{Petrov1999, Petrov2000}, and Dirac-based B-spline R-matrix \cite{Zatsarinny2010} treatments.
These treatments are usually satisfactory just above threshold or away from threshold, predicting the ``general" behavior of the observed data.

The measurement of cross sections just above threshold has been limited to only a handful of efforts, spanning several decades, \cite{Hudson1965, Marr1968, Ditchburn1953, Sandner1981}, in part because the absolute measurement is hampered by the existence of, \textit{ inter alia}, alkali metal dimers with large photoabsorption cross sections that
mask the atomic contributions \cite{Granneman1975, CET1979}. 
Sandner et al.\ \cite{Sandner1981} have overcome such complications  in K by using a time-of-flight technique to separate the atomic K signal. Their high-accuracy results brought \emph{qualitative} agreement between experiment and theory but a significant discrepancy in the width of the Cooper minimum remains.

Two recent calculations \cite{Petrov1999, Zatsarinny2010} have revisited the K($4s$) photoionization minimum and three experimental papers \cite{Amin2008,Yar2013, Kalyar2016} have measured photoionization cross sections from excited ($4p,5d,7s$) potassium states.
The theoretical works reproduce the K($4s$) position and depth of the minimum fairly accurately, but the shape is wider than the accurate experimental determination of Sandner et al.\ \cite{Sandner1981}, considered as the most accurate reference point.
There  remains elusive a fully \textit{ab initio} calculation, without using any fitting parameters, of the K($4s$) photoionization cross section near threshold, that agrees with high accuracy experimental results.

In view of the current situation, we have revisited our earlier preliminary report on this system \cite{CET2003} and present here the refined results. Our data reproduce the Sandner et al.\ data exactly, especially the minimum width. Furthermore, they seem to indicate that the Hudson and Carter \cite{Hudson1965} measurements are more accurate than those of Marr and Creek \cite{Marr1968} for energies above the minimum.

Our present treatment resolves this width discrepancy through a careful treatment of the spin-orbit interaction and core-polarization and it shows that the use of a dynamic, rather than static, core-polarizability reproduces the observed \cite{Hudson1965} rise of the cross section just below the threshold for the lowest core-excited state.

\section{Method of Calculation}

A semi-empirical method was developed by this author \cite{CET1984a} to calculate accurate transition matrix elements for alkali-like and helium-like systems for which no core excitation or channel mixing is present. The method employs a self-consistent Hartree-Kohn-Sham potential to describe the atomic field and utilizes the experimental energy levels as input, supplemented by {\it ab initio} core-polarizabilities. The method has been successful in predicting accurate oscillator strengths and excited state lifetimes for a variety of atomic systems, including Li, He, Be$^{+}$, Na, K, Ca$^{+}$, Cu, Rb, Zn$^{+}$, Cd$ ^{+}$, Cs, Ba$^{+}$, and Fr lifetimes (e.g., see  \cite{CET1984b}, \cite{Curtis1993}, \cite{CET1996} and references therein). The general approach is refined here to account for more salient features of the transitions at hand.

\begin{widetext}
The wave functions are obtained by solving the Schr\"odinger equation
\begin{equation}
\left[ \frac{d^2}{dr^2}-V(r)-\frac{l(l+1)}{r^2}+E_{nl}\right]
P_{nl}(r)=0.
\end{equation}
The central potential
\begin{equation}
V(r)=V_{{\rm HKS}}(r)+V_{{\rm pol}}(r)+V_{{\rm so}}(r)=V_{{\rm m}}(r)+V_{%
{\rm so}}(r)
\end{equation}
consists of three terms: $V_{\rm HKS}(r)$, a Hartree-Kohn-Sham-type \cite{Desclaux1969} self-consistent field term,

\begin{equation}
V_{\rm pol}(r)=-\frac 12\frac{\alpha _d}{r^4}\left\{ 1-{\rm exp}
\left[-(r/r_c)^6\right] \right\}                             
\\
 -\frac 12\frac{\alpha _q}{r^6}\left\{ 1-{\rm exp}%
\left[ -(r/r_c)^{10}\right] \right\}
\end{equation}
a core-polarization term, and
\begin{equation}
V_{{\rm so}}(r)=-\frac 12\alpha ^2\left\{ 1+\frac{\alpha ^2}%
4[E-V(r)]\right\} ^{-2}\frac 1r\frac{dV_{\rm m}(r)}{dr}\vec L\cdot \vec S
\end{equation}
a spin-orbit interaction, Pauli approximation term. Here $\alpha_d$ and $ \alpha_q$ are the dipole and quadrupole polarizabilities of the core \cite{Johnson1981}, $r_c$ is a cut-off distance and $\alpha$ is the fine-structure constant.

The necessary radial matrix elements were calculated using the modified dipole operator expression similar to the one used by Norcross \cite{Norcross1973},

\begin{equation}
R(nl,n'l'j')=\left< n'l'j'\left| r\left\{ \left[ 1-\frac{\alpha _d}{r^3}\left( 1-\frac
12\left[ {\rm exp}\left\{ -(r/r_{cl})^3\right\} +{\rm exp}\left\{
-(r/r_{cl^\prime })^3\right\} \right] \right) \right] \right\}
\right| nlj\right>.
\end{equation}

The cutoff distances $r_{cl}$ are taken to be equal to the values used in the polarization potential $V_{pol}$ needed to reproduce the lowest experimental energy for each symmetry; they are different for each value of $l$.  They are the only adjustable parameters in this approach.

The photoabsorption cross section is obtained using \cite{Sobelman1992}
\begin{equation}
\sigma(nlj\rightarrow n'l'j')=\frac43 \pi^2a_0^2\alpha
(\epsilon_{n'l'j'}-\epsilon_{nlj})\frac{1}{2j+1} S(nlj,n'l'j')
\end{equation}
where energy is given in Rydbergs, and the absorption oscillator strength $f$ is 
\begin{equation}
f(nlj\rightarrow n'l'j')=\frac23(\epsilon_{n'l'j'}-\epsilon_{nlj})
\frac{1}{2j+1}S(nlj,n'l'j')
\end{equation}
The line strength $S(nlj,n'l'j')$ is the same as the reduced dipole matrix element and in the case of alkali metal atoms is given by
\begin{equation}
S(nlj,n'l'j')=\text{max}(l,l')(2j+1)(2j'+1)
\begin{Bmatrix}  
l &{\ \ \ \ } l' &{\ \ \ \ } 1 \\
j' &{\ \ \ \ } j &{\ \ \ \ } \frac{1}{2} \\
\end{Bmatrix}
^2
R(nlj,n'l'j')^2.
\end{equation}

When the final state is in the continuum, $n'l'j'$ is replaced by $\epsilon l'j'$ and the total photoionization cross section of an initial state $|nlj>$ to the continuum comprises from
the sum of the partial cross sections for a photon energy $E=\epsilon-\epsilon_{nlj}$:
\begin{equation}
\sigma_{nlj}(E)=\frac43 \pi^2a_0^2\alpha (\epsilon-\epsilon_{nlj})
\sum_{l'j'}\text{max}(l,l')(2j'+1)
\begin{Bmatrix}  
l &{\ \ \ \ } l' &{\ \ \ \ } 1 \\
j' &{\ \ \ \ } j &{\ \ \ \ } \frac{1}{2} \\
\end{Bmatrix}
^2
R(nlj,\epsilon l'j')^2.
\end{equation}
\end{widetext}

The continuum photoionization cross section $\sigma(nlj$$\rightarrow $$n'l'j')$ joins smoothly at the energy threshold with the oscillator strength distribution for discrete states using the formula
\begin{equation}
\sigma(nlj\rightarrow{}n'l'j')=\frac{2\pi^2\alpha\hbar^2}{m} \frac{df}{dE}\\
\end{equation}
where
\begin{equation}
\frac{df}{dE}\equiv \frac{df_{n'l'j'}}{d\epsilon_{n'l'j'}} =\frac{1}{2}(n_{j'}^*)^3
f(nlj\rightarrow n'l'j').
\end{equation}

\section{Results and Discussion}

\subsection{ K $4s$ state}

\begin{figure}[h]\centering\small\label{K-4s_cs_m}
\includegraphics[width=8.6cm]
{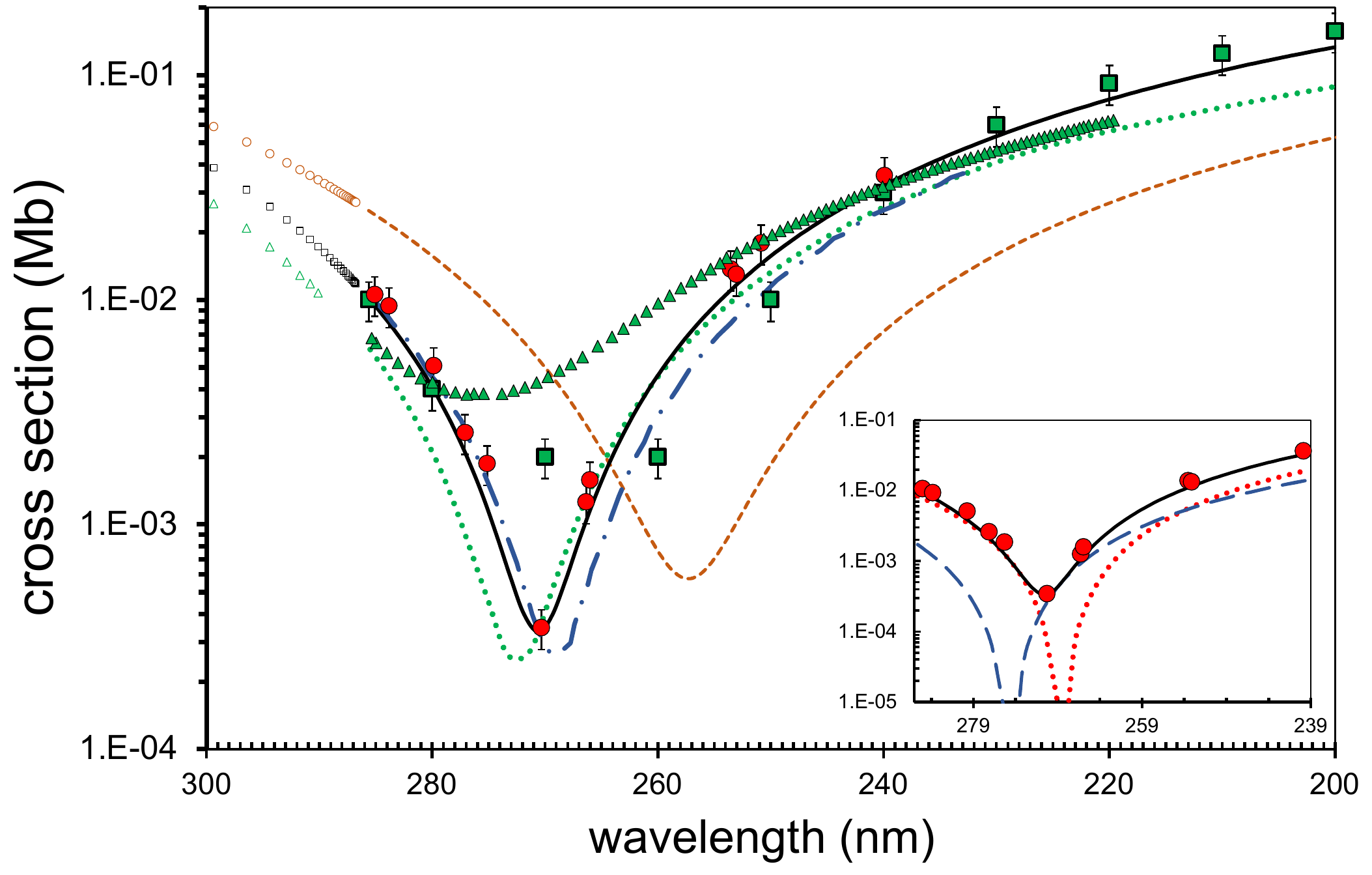}
\caption{\label{fig: K-4s_cs_m} (Color online) K($4s$) photoabsorption cross section around the first threshold.
Experimental: $\mdblkcircle$, Sandner et al.\ \cite{Sandner1981}, $\mdwhtsquare$, Hudson and Carter \cite{Hudson1965}, $\bigtriangleup$, Marr and Creek \cite{Marr1968}.
Theoretical: dots (green), Weisheit \cite{Weisheit1972}; dashed (orange), using the potentials of Marinescu et al.\ \cite{Marinescu1994}; dash-dot (blue), Zatsarinny and Tayal \cite{Zatsarinny2010}; solid line, present results.  The respective data below the threshold represent the optical oscillator strengths. \textit{Insert:} Partial-wave contributions to the K($4s$) photoionization cross section.  $\mdblkcircle$, Sandner et al.\ \cite{Sandner1981}; dots (red), $4s$$\rightarrow$$\epsilon p_{1/2}$ cross section; dashes (blue),  $4s$$\rightarrow$$\epsilon p_{3/2}$ cross section; solid line, total cross section.}
\end{figure}

Figure \ref{fig: K-4s_cs_m} displays the results of our calculations for the K $4s$ photoabsorption, along with those using the published semiempirical potentials by Weisheit \cite{Weisheit1972} and by Marinescu et al.\ \cite{Marinescu1994}, and they are compared with the three available experiments near threshold\cite{Hudson1965, Sandner1981}.
The calculations extend from photoabsorption to high Rydberg states below the first ionization threshold, to continuum states well above the well-known minimum above threshold.
All three calculations reproduce the general shape of the minimum but vary in the prediction of its location. Our results reproduce best the data by Sandner et al.\ \cite{Sandner1981}.
The older data by Hudson and Carter \cite{Hudson1965} reproduce the ``wings'' well, but are not accurate around the the bottom of the minimum. The data Marr and Creek\cite{Marr1968} display a shallow minimum and grow slower with increasing energy.

The insert in Figure \ref{fig: K-4s_cs_m} shows the contributions of the $\epsilon p_{1/2}$ and $\epsilon p_{3/2}$ separately and summed. The two corresponding minima do reach zero, but being separated in energy result in a finite-value total minimum.

It has been pointed in the literature, e.g. see \cite{CET1984a,Norcross1973}, that for the transition matrix element a more physical treatment would be to use the dynamic core polarizability $\alpha _d (\omega )$ where $\hbar \omega$ is the photon energy, instead of the static value $\alpha _d (0) \equiv \alpha _d $.  To a first approximation, the two quantities are related as
\begin{equation}
\alpha (\omega)\approx \alpha (0)/[1-(\hbar\omega/\Delta
E_{rc})^2],
\end{equation}
where $\Delta E_{rc}$ is the resonance energy of the core.  The approximation is valid for frequencies less than this energy.

\begin{figure}[h]\centering\small\label{Fig2}
\includegraphics[width=8.5cm]{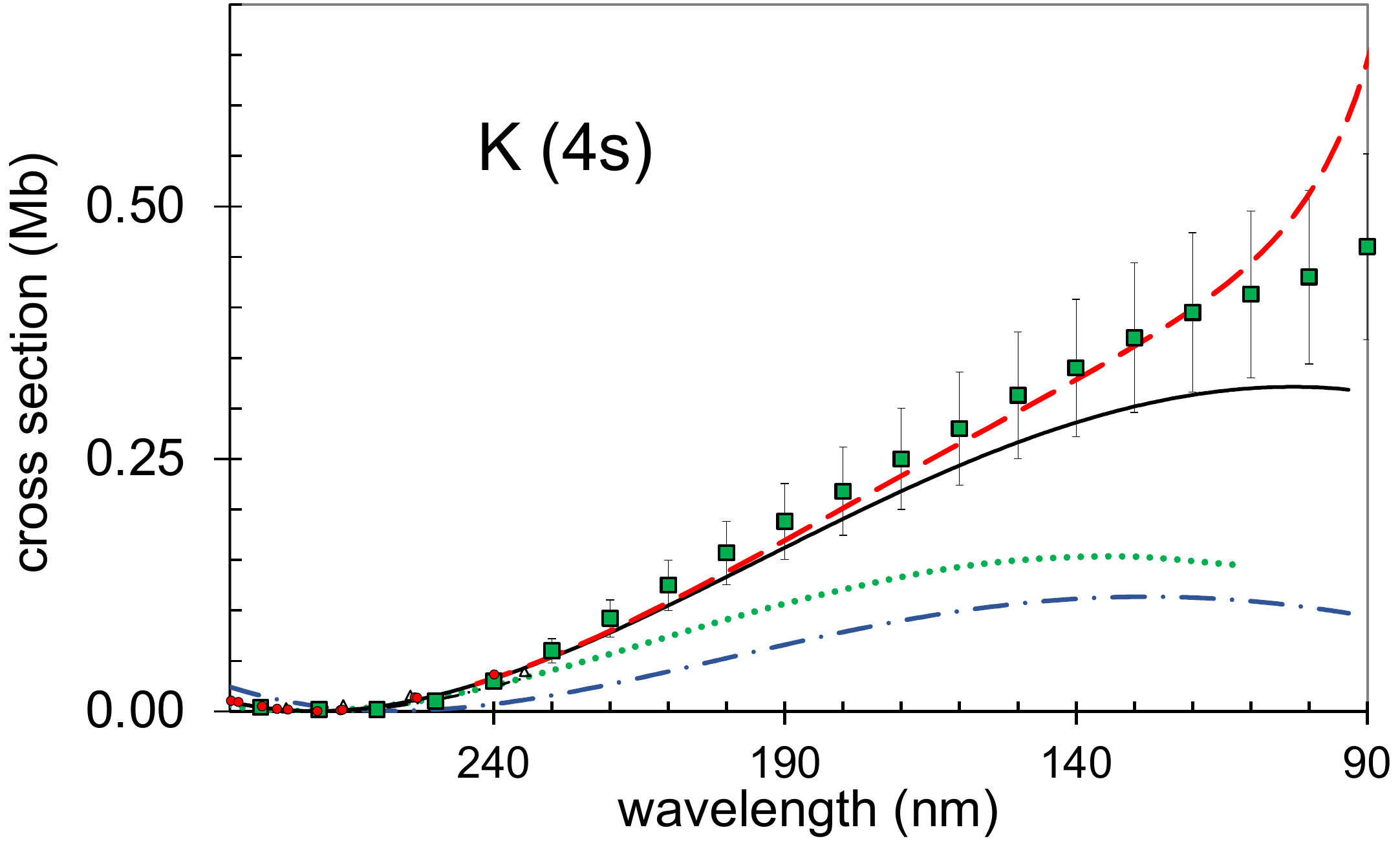}
\caption{\label{fig: K-4s_cs_d_c}(Color online)  K ($4s$) photoionization cross section below the core resonance threshold.
Experimental: $\mdwhtsquare$, Hudson and Carter\cite{Hudson1965}.
Theoretical: dots (green), Weisheit\cite{Weisheit1972}; dash-dot (blue), Zatsarinny and Tayal\cite{Zatsarinny2010}; solid line, present results with $a_d(0)$; dashed line (red), present results with $a_d(\omega)$. }
\end{figure}

Figure \ref{fig: K-4s_cs_d_c} shows that the use of the dynamic polarizability reproduces  the upward trend of the experimental data of Hudson and Carter \cite{Hudson1965} at higher energies as one approaches the core excitation threshold of about 18.8 eV.

\subsection{K $7s,$ $4p$, and $5d$ states}
The work of Amin et al.\cite{Amin2008} presented data on the $4p$ photoionization that seem to be in fair agreement with the experimental and theoretical results of Petrov et al.\cite{Petrov2000}.  However, their measured cross sections \cite{Amin2008} for $7s$ and $5d$ photoionization strongly disagree with the theoretical calculations of Zatsarinny and
Tayal\cite{Zatsarinny2010}. Following the publication of the latter results, Yar et al.\cite{Yar2013} remeasured the $7s$ photoionization cross section using a different experimental approach and their results reproduce closely the calculations of Zatsarinny and Tayal \cite{Zatsarinny2010} near the predicted Cooper minimum around photoelectron energy 1.2 eV.

\begin{figure}[h]\centering\small\label{Fig3}
\includegraphics[width=8.6cm]{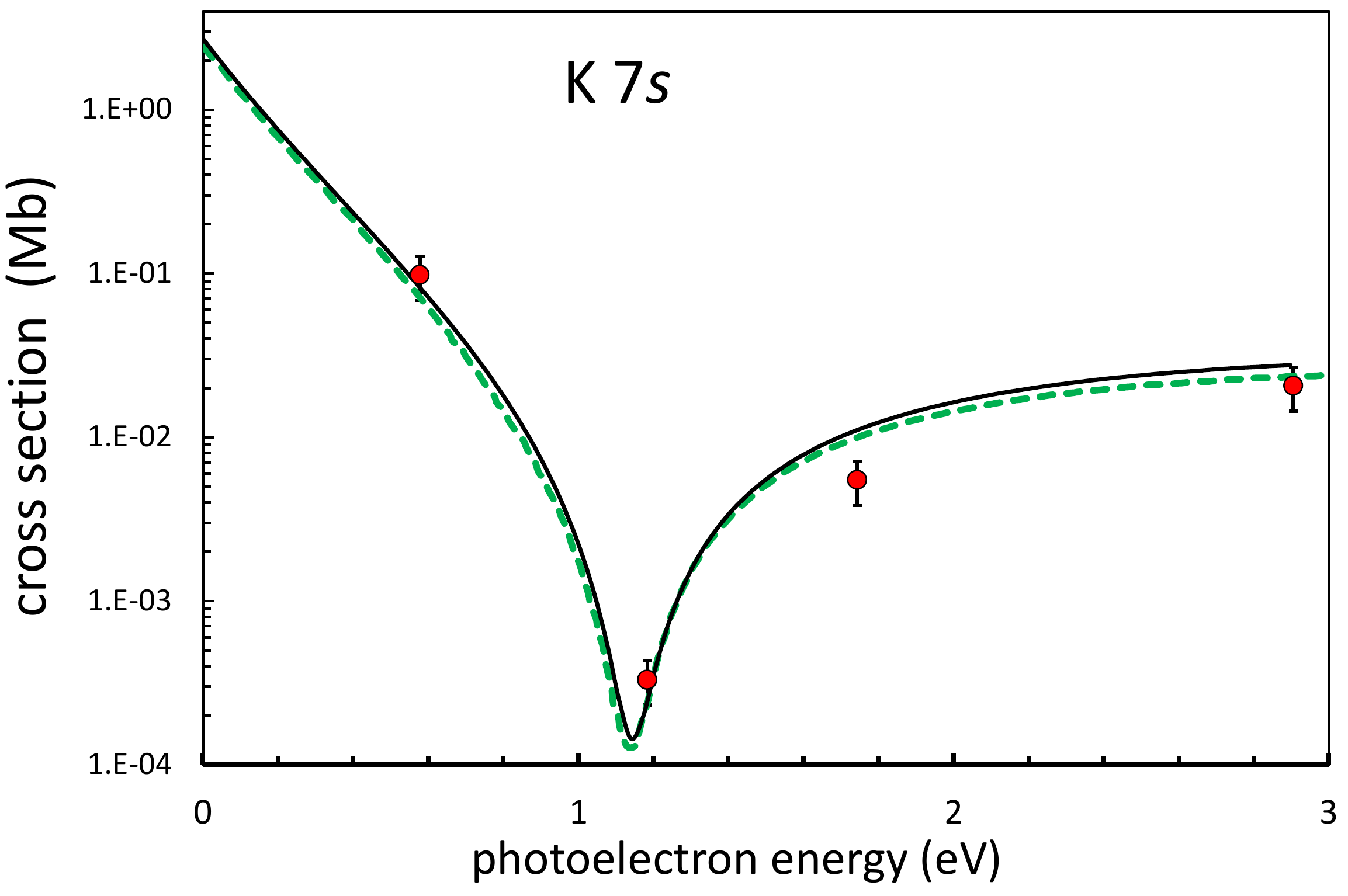}
\caption{\label{fig: K-7s_photo}  
K $7s$ photoionization cross section.  $\mdblkcircle$, Yar et al.\ \cite{Yar2013}; dashed (green), Zatsarinny and Tayal \cite{Zatsarinny2010}; continuous line, present results.}
\end{figure}

Using our approach we calculated the K ($7s$) photoinization cross section from threshold through the energy range of the Cooper minimum.
Our results (Figure \ref{fig: K-7s_photo}) are in excellent agreement with the measurements of Yar et al.\ \cite{Yar2013} and the calculations of Zatsarinny and Tayal \cite{Zatsarinny2010}. As with the case of the $4s$ photoionization, our minimum ``width'' is narrower than that of Ref. \cite{Zatsarinny2010}. 

As a routine application of our approach we also calculated the photoabsorption from the $4p$ and $5d$ states. Our results for $5d$ are essentially identical to those of Zatsarinny and Tayal \cite{Zatsarinny2010} and disagree with the data of Ref.\ \cite{Amin2008} by a factor of ten. Thus, we see no need to present a redundant graph of our results on this state.

In the case of $4p$ we have for comparison four different measurements \cite{Burkhardt1988, Amin2008, Petrov2000,Kalyar2016} and the relatively recent calculations of Petrov et al.\cite{Petrov2000} and Zatsarinny and Tayal\cite{Zatsarinny2010}. We consider the
\begin{figure}[b]\centering\small\label{Fig4}
\includegraphics[width=8.0cm]{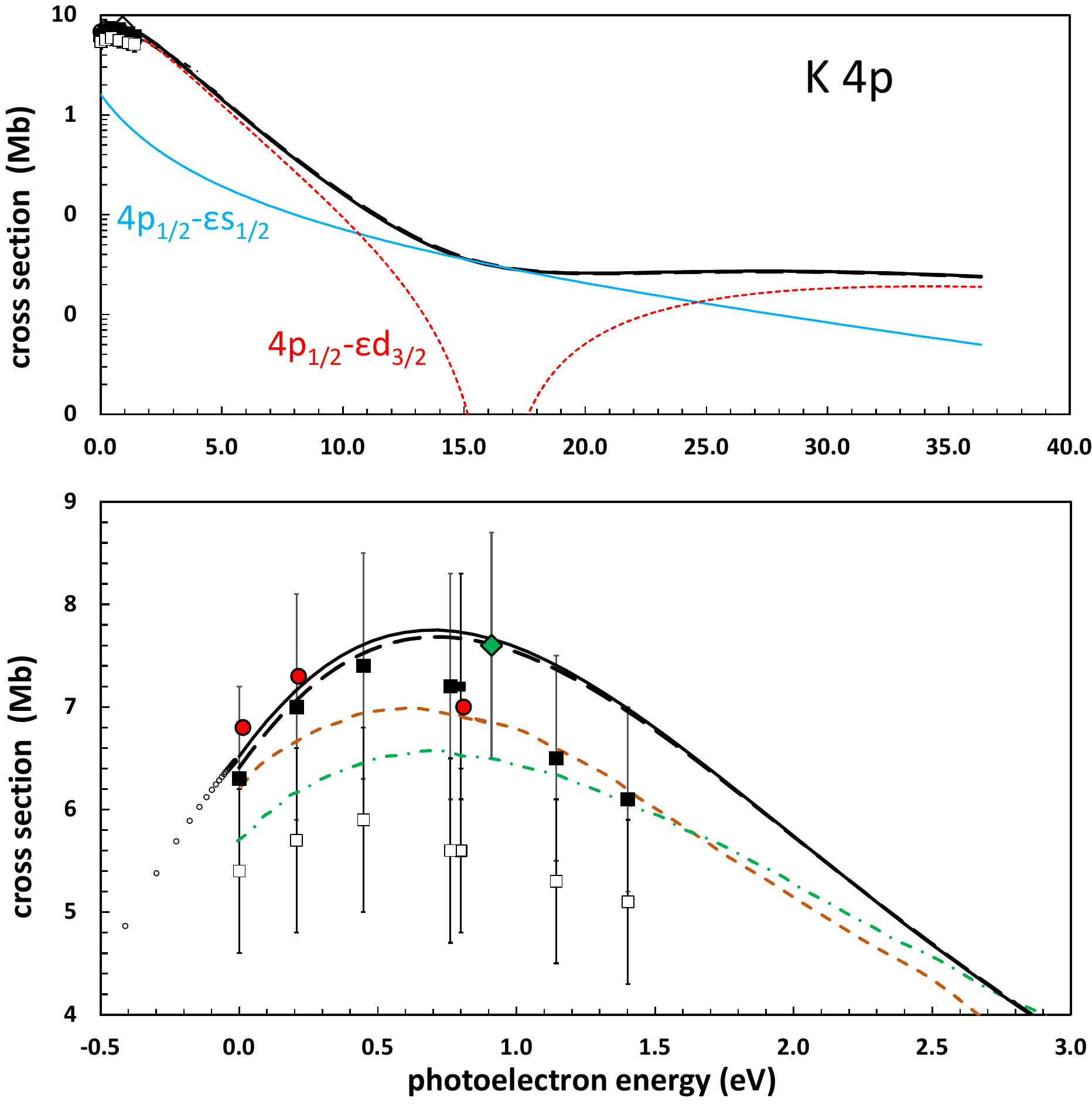}
\caption{\label{fig: K-4p_cs}  (Color online) K $4p$ photoionization cross sections.  \textit{Top}: Thick (black) line, $4p_{3/2}$, present; thin solid line (cyan), $4p_{1/2}$$\rightarrow$$\epsilon s_{1/2}$, present; 
dotted line (red), $4p_{1/2}$$\rightarrow$$\epsilon d_{3/2}$, present.  \textit{Bottom}: (enlarged view near threshold) 
Experimental: $\mdwhtcircle$ (red), Petrov et al.\cite{Petrov2000}; $\mdwhtdiamond$ (green), Burkhardt et al.\cite{Burkhardt1988}; $\mdblksquare$, $4p_{3/2}$, Kalyar et al.\cite{Kalyar2016}; $\mdwhtsquare$, $4p_{1/2}$, Kalyar et al.\cite{Kalyar2016}.
Theoretical: dashed (red), Petrov et al.\cite{Petrov2000}; dash-dot (green), Zatsarinny and Tayal\cite{Zatsarinny2010}; solid line, $4p_{3/2}$, present results; long-dashed line, $4p_{1/2}$, present results.  The points below threshold represent the Rydberg state photoabsorption.}
\end{figure}
\noindent
measurements of Burkhardt et al. \cite{Burkhardt1988} as an important reference point, since they claim to be absolute values and have been withstood the test of comparison over the years for both K(4$p$) and Na(3$p$) photoionization. The behavior of the cross section is smooth and we found a very shallow minimum considerably higher up, around 20 eV, caused by a Cooper minimum in the $\sigma(4p-\epsilon s)$ at 16.3 eV.. Our calculations (Figure \ref{fig: K-4p_cs}) are in excellent agreement with the experimental results of Burkhardt et al.\cite{Burkhardt1988} and Petrov et al.\cite{Petrov2000} and in agreement with the $4p_{3/2}$ measurement of Amin et al.\cite{Amin2008} and Kalyar et al.\cite{Kalyar2016}, but not the $4p_{1/2}$ state of the latter two works. Our results are also in good agreement with the theoretical ones of Petrov et al.\  but less so with those of Zatsarinny and Tayal, who see a few percent difference between the $4p_{1/2}$ and $4p_{3/2}$ cross sections. The difference we find between the values for $np_{3/2}$ and $np_{1/2}$ is relatively small, 1.6\% at threshold.

Since the work of Kalyar et al.\cite{Kalyar2016} presents data and discussion of the $4p$ to $nd$ absorption oscillator strengths, we examined the transitions to $nd$ and $ns$ states with $n=4-80$. Our results are indicated in Figure \ref{fig: K-4p_cs} for a few discreet
\begin{figure}[h]\centering\small\label{Fig5}
\includegraphics[width=8.6cm]{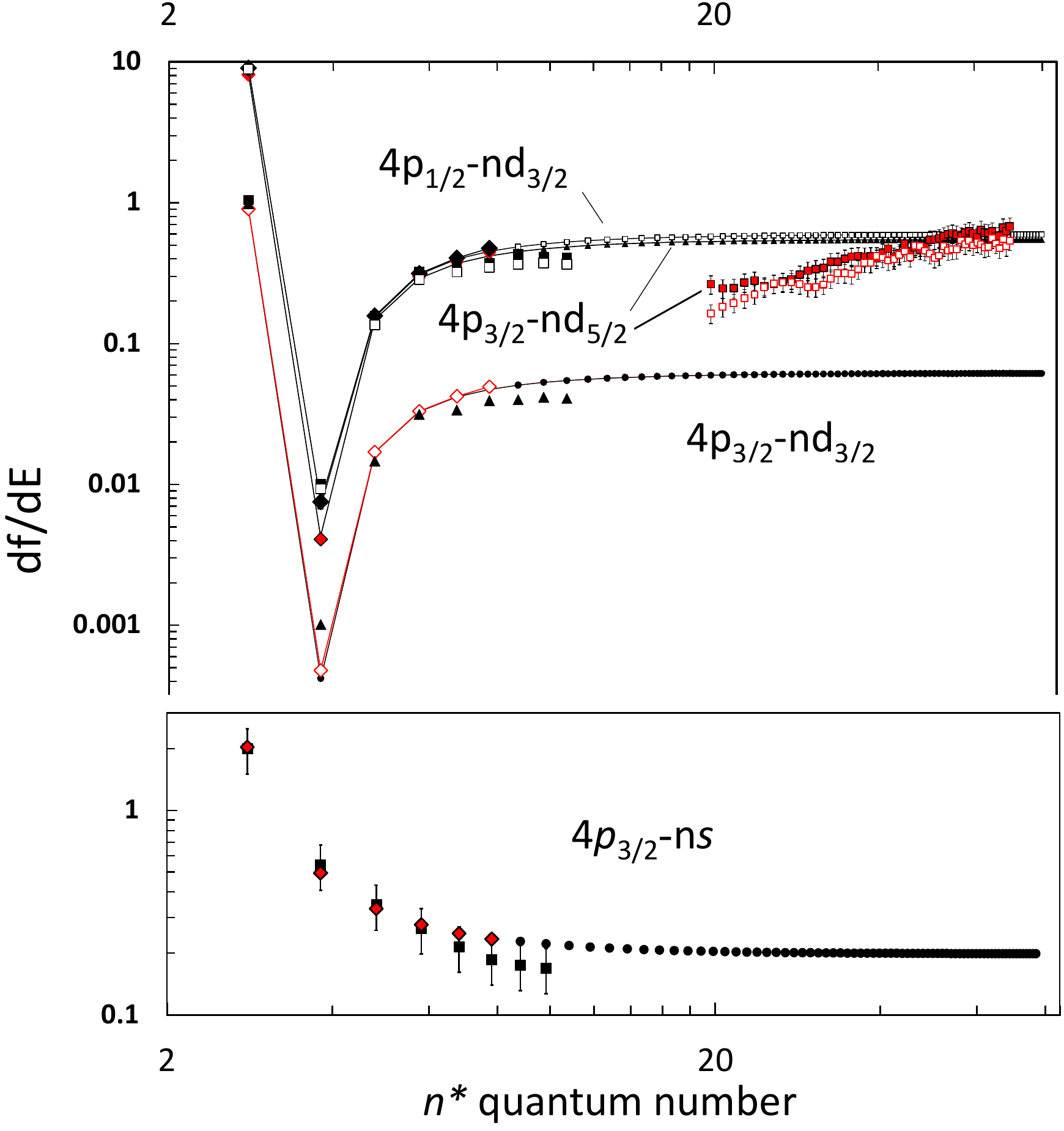}
\caption{\label{fig: K-4p-nd_df_dE}   (Color online)  \textit{Top}: K $4p-nd_{3/2,1/2}$ absorption oscillator strengths. 
Experimental:  $\mdblksquare$ (red), $4p_{3/2}$, Kalyar et al.\cite{Kalyar2016}; $\mdwhtsquare$ (red), $4p_{1/2}$, Kalyar et al.\cite{Kalyar2016}; for $n < 11$, $\mdblksquare$ and $\mdwhtsquare$, data taken from Wiese et al.\cite{Wiese1969};   
Theoretical: $\mdblkdiamond$, $\mdwhtdiamond$, using the matrix elements of Ref.\ \cite{Safronova2013}; $\mdblkcircle$, present $4p_{3/2}$$\rightarrow$$nd_{5/2}$; $\mdwhtcircle$, present $4p_{1/2}$$\rightarrow$$nd_{3/2}$; $\bigtriangleup$, present $4p_{3/2}$$\rightarrow$$nd_{3/2}$. 
 \textit{Bottom}: K $4p$$\rightarrow$$ns$ oscillator strengths. Experimental:  $\mdblksquare$, Wiese et al.\cite{Wiese1969}.  
Theoretical: $\mdwhtdiamond$, using the matrix elements of Ref.\ \cite{Safronova2013}; $\mdblkcircle$, present $4p_{1/2}$.} 
\end{figure}
\noindent
transitions as values for ``negative energy,'' to show the smooth transition between discrete and continuum photoabsorption. Then they are presented, in Figures \ref{fig: K-4p-nd_df_dE}, in the form of $df/dE=f n^{*3}/2$, where $n^*$ is the effective quantum number, and compared with the data of Kalyar et al.\ \cite{Kalyar2016}, along with the very early tabulation of Wiese et al.\ \cite{Wiese1969} as a standard reference. From Figure \ref{fig: K-4p-nd_df_dE} it is seen that our data have a smooth transition from $\text{n}=4$ to $\text{n}=80$, in very good agreement with the data of Wiese et al.\ up to $n=11$, and reaching the asymptotic bahavior of $1/n^{*3}$. 
The data of Kalyar at al.\cite{Kalyar2016} agree with ours above $n=35$, although they are not smooth; for lower n, however, they deviate from the reasonably expected
$1/n^{*3}$ behavior. It is interesting that the analogous graph of the latter paper (their Fig.\ 8) shows two clearly separated lines, whereas that does not correspond to their published values for the oscillator strengths $4p_{3/2}$$\rightarrow$$\text{n}d_{3/2, 5/2}$ and $4p_{1/2}$$\rightarrow$$\text{n}d_{3/2}$.
\begin{table}[t]\small
\centering\label{Table1}
\caption{Absolute values of reduced dipole matrix elements. CET, this work; SSC, Ref.\ \cite{Safronova2013}; NSSS, Ref.\ \cite{Nandy2012};
JTM,  Ref.\ \cite{Jiang2013}.}
\begin{tabular}{rc *{4}c}
\toprule\toprule 
\multicolumn{2}{c}{Transition}&{\ \ \ \ } 
{CET}     &{\ \ \ \ }{ SSC}	     &{\ \ \ \ } {NSSS}        &{\ \ \ \ }{JTM}	   \\
\midrule
\multicolumn{3}{l}{$4p_{1/2}$-$\text{n}d_{3/2}$}\\
\midrule
n=&{\ \ \ \ }3 &{\ \ \ \ } 7.902	 &{\ \ \ \ } 7.979(35)	      &{\ \ \ \ } 7.988(40) &{\ \ \ \ } 7.966\\
   &{\ \ \ \ }4&{\ \ \ \ } 0.108	 &{\ \ \ \ } 0.112(5)	    &{\ \ \ \ } 0.220(5)  &{\ \ \ \ } 0.140	 \\
   &{\ \ \ \ }5&{\ \ \ \ } 0.328	 &{\ \ \ \ } 0.333(5)		 	&{\ \ \ \ } 0.264(5)\\
   &{\ \ \ \ }6&{\ \ \ \ } 0.337      &{\ \ \ \ }0.341(5)		&{\ \ \ \ } 0.293(5)\\
   &{\ \ \ \ }7&{\ \ \ \ } 0.294	 &{\ \ \ \ } 0.298(5)		 	&{\ \ \ \ }  0.261(4)\\
   &{\ \ \ \ }8&{\ \ \ \ } 0.251      &{\ \ \ \ }0.254(4)	  &{\ \ \ \ } 0.221(4)\\
   &{\ \ \ \ }9&{\ \ \ \ } 0.215	 &{\ \ \ \ } 0.218(3)		\\
\midrule 
\multicolumn{3}{l}{$4p_{3/2}$-$\text{n}d_{3/2}$}\\
\midrule
n=
&{\ \ \ \ }3   &{\ \ \ \ }3.544	&{\ \ \ \ }3.578(16)	&{\ \ \ \ }3.583(20)	&{\ \ \ \ }3.573\\
&{\ \ \ \ }4	&{\ \ \ \ }0.039	&{\ \ \ \ }0.040(6)	&{\ \ \ \ }0.088(5)	&{\ \ \ \ }0.053\\
&{\ \ \ \ }5  &{\ \ \ \ }0.153	&{\ \ \ \ }0.155(2)	&{\ \ \ \ }0.124(5)	\\
&{\ \ \ \ }6	&{\ \ \ \ }0.155	&{\ \ \ \ }0.157(2)	&{\ \ \ \ }0.135(5)	\\
&{\ \ \ \ }7	&{\ \ \ \ }0.135	&{\ \ \ \ }0.136(2)	&{\ \ \ \ }0.119(3)	\\
&{\ \ \ \ }8	&{\ \ \ \ }0.115	&{\ \ \ \ }0.116(2)	&{\ \ \ \ }0.101(3)	\\ 
&{\ \ \ \ }9&{\ \ \ \ }0.098&{\ \ \ \ }	0.100(2)		\\
\midrule
\multicolumn{3}{l}{$4p_{3/2}$-$\text{n}d_{5/2}$}\\
\midrule
n=
&{\ \ \ \ }3&{\ \ \ \ }10.635		&{\ \ \ \ }10.734(47)		&{\ \ \ \ }10.75(5)		&{\ \ \ \ }10.719\\
&{\ \ \ \ }4&{\ \ \ \ }0.118		&{\ \ \ \ }0.117(15)		&{\ \ \ \ }0.260(5)		&{\ \ \ \ }0.155\\
&{\ \ \ \ }5	&{\ \ \ \ }0.460		&{\ \ \ \ }0.467(6)		&{\ \ \ \ }0.374(5)	\\
&{\ \ \ \ }6	&{\ \ \ \ }0.465		&{\ \ \ \ }0.471(7)		&{\ \ \ \ }0.404(5)	\\
&{\ \ \ \ }7	&{\ \ \ \ }0.404		&{\ \ \ \ }0.409(7)		&{\ \ \ \ }0.356(5)	\\
&{\ \ \ \ }8	&{\ \ \ \ }0.345		&{\ \ \ \ }0.349(5)		&{\ \ \ \ }0.286(5)	\\
&{\ \ \ \ }9	&{\ \ \ \ }0.295		&{\ \ \ \ }0.299(4)		\\
\midrule
\multicolumn{3}{l}{$4p_{1/2}$-$\text{n}s$}\\
\midrule
n=
&{\ \ \ \ }5   &{\ \ \ \ }3.883	&{\ \ \ \ }3.885(8)		&{\ \ \ \ }3.876(10)		&{\ \ \ \ }3.888\\
&{\ \ \ \ }6	&{\ \ \ \ }0.910		&{\ \ \ \ }0.903(4)		&{\ \ \ \ }0.909(10)	\\
&{\ \ \ \ }7	&{\ \ \ \ }0.481		&{\ \ \ \ }0.476(2)		&{\ \ \ \ }0.479(5)		\\
&{\ \ \ \ }8	&{\ \ \ \ }0.317		&{\ \ \ \ }0.314(2)		&{\ \ \ \ }0.316(5)	\\
&{\ \ \ \ }9	&{\ \ \ \ }0.233		&{\ \ \ \ }0.230(1)		&{\ \ \ \ }0.225(3)	\\
&{\ \ \ \ }10	&{\ \ \ \ }0.181		&{\ \ \ \ }0.1791(9)		&{\ \ \ \ }0.171(3)	\\
&{\ \ \ \ }11	&{\ \ \ \ }0.147		&{\ \ \ \ }0.1452(8)		\\
\midrule
\multicolumn{3}{l}{$4p_{3/2}$-$\text{n}s$}\\
\midrule
n=
&{\ \ \ \ }5&{\ \ \ \ }5.528		&{\ \ \ \ }5.54(1)		&{\ \ \ \ }5.52(2)		&{\ \ \ \ }5.538\\
&{\ \ \ \ }6	&{\ \ \ \ }1.286		&{\ \ \ \ }1.279(5)		&{\ \ \ \ }1.287(10)	\\
&{\ \ \ \ }7	&{\ \ \ \ }0.678		&{\ \ \ \ }0.673(3)		&{\ \ \ \ }0.677(6)	\\
&{\ \ \ \ }8	&{\ \ \ \ }0.448		&{\ \ \ \ }0.444(2)		&{\ \ \ \ }0.447(5)	\\
&{\ \ \ \ }9	&{\ \ \ \ }0.328		&{\ \ \ \ }0.325(2)		&{\ \ \ \ }0.317(5)	\\
&{\ \ \ \ }10	&{\ \ \ \ }0.255		&{\ \ \ \ }0.253(1)		&{\ \ \ \ }0.242(5)	\\
&{\ \ \ \ }11	&{\ \ \ \ }0.207		&{\ \ \ \ }0.205(1)		\\
\bottomrule
\end{tabular}
\end{table}

Our $4p-ns$ oscillator strengths are also presented in Figure \ref{fig: K-4p-nd_df_dE} for the completeness. The large-$n$ behavior is reaching a $1/n^{*3}$ dependence after about $n=10$. Here, too, we find good agreement with the data of Wiese et al.\cite{Wiese1969} for the values of $n$ where data are available.

Since the Wiese et al.\cite{Wiese1969} data are over 50 years old and carry a 25-50\% estimated error, we compared our $4p_j - \text{n}d_{j'}$ and $4p_j -\text{n}s$ reduced electric-dipole
matrix elements (i.e. the $\sqrt{S(nlj,n'l'j')}$ in Eq.(8)) to those published by Safronova et al.\cite{Safronova2013} and considered to be of high accuracy. The agreement between the two sets is within 1\% for all quoted matrix elements (see Table 1). Additional comparison is also made in Table 1 with Jiang et al.\cite{Jiang2013} and Nandy et al.\cite{Nandy2012}; here the agreement is not as consistently good. These comparisons give us added confidence in the overall accuracy of our calculations and the validity of the presented arguments.  The matrix elements from Safronova were converted to oscillator strengths and are also presented on Figures \ref{fig: K-4p-nd_df_dE}.  The agreement with our values is excellent.  Thus we can assume that our oscillator strengths over the entire Rydberg series can be considered as a revised reference point, replacing the data of Wiese et al.\cite{Wiese1969}.
Table II gives the recommended $4p_j-\text{n}s,\text{n}d$ oscillator strengths through $\text{n}=80.$



\begin{longtable}[h]{c|rrr|rr}
\caption{
Oscillator strengths  $4p_{3/2}$-$\text{n}d,\text{n}s$  and $4p_{1/2}$-$\text{n}d, \text{n}s$}\label{}  \\
\midrule \midrule
\multicolumn{1}{c}{} & \multicolumn{3}{l}{$4p_{3/2}$→} & \multicolumn{2}{l}{$4p_{1/2}$→} \\
\midrule
{n} &$\text{n}d_{5/2}${\ \ \ } &$\text{n}d_{3/2}${\ \ \ }  &$\text{n}s_{1/2}${\ \ \ }  &$\text{n}d_{3/2}${\ \ \ }  &$\text{n}s_{1/2}${\ \ \ } \\
\midrule 
\endfirsthead

\midrule \midrule
\multicolumn{1}{c}{} & \multicolumn{3}{l}{$4p_{3/2}$→} & \multicolumn{2}{l}{$4p_{1/2}$→} \\
\midrule
{$n$} &$nd_{5/2}${\ \ \ } &$\text{n}d_{3/2}${\ \ \ }  &$\text{n}s_{1/2}${\ \ \ }  &$\text{n}d_{3/2}${\ \ \ }  &$\text{n}s_{1/2}${\ \ \ } \\
\midrule 
\endhead
\midrule 
\endfoot
\midrule
\endlastfoot
  3	&{\ \ \ }	7.31E-01	&{\ \ \ }	8.12E-02	&{\ \ \ }		&{\ \ \ }	8.07E-01	&{\ \ \ }		\\
  4	&{\ \ \ }	1.60E-04	&{\ \ \ }	1.53E-05	&{\ \ \ }	6.69E-01	&{\ \ \ }	2.57E-04	&{\ \ \ }	3.34E-01	\\
  5	&{\ \ \ }	2.73E-03	&{\ \ \ }	3.12E-04	&{\ \ \ }	1.86E-01	&{\ \ \ }	2.85E-03	&{\ \ \ }	1.83E-01	\\
  6	&{\ \ \ }	3.04E-03	&{\ \ \ }	3.44E-04	&{\ \ \ }	1.81E-02	&{\ \ \ }	3.25E-03	&{\ \ \ }	1.81E-02	\\
  7	&{\ \ \ }	2.42E-03	&{\ \ \ }	2.73E-04	&{\ \ \ }	6.03E-03	&{\ \ \ }	2.60E-03	&{\ \ \ }	6.03E-03	\\
  8	&{\ \ \ }	1.81E-03	&{\ \ \ }	2.04E-04	&{\ \ \ }	2.85E-03	&{\ \ \ }	1.95E-03	&{\ \ \ }	2.86E-03	\\
  9	&{\ \ \ }	1.35E-03	&{\ \ \ }	1.52E-04	&{\ \ \ }	1.60E-03	&{\ \ \ }	1.46E-03	&{\ \ \ }	1.61E-03	\\
10	&{\ \ \ }	1.02E-03	&{\ \ \ }	1.15E-04	&{\ \ \ }	9.99E-04	&{\ \ \ }	1.11E-03	&{\ \ \ }	1.00E-03	\\
11	&{\ \ \ }	7.86E-04	&{\ \ \ }	8.84E-05	&{\ \ \ }	6.69E-04	&{\ \ \ }	8.51E-04	&{\ \ \ }	6.71E-04	\\
12	&{\ \ \ }	6.15E-04	&{\ \ \ }	6.92E-05	&{\ \ \ }	4.71E-04	&{\ \ \ }	6.67E-04	&{\ \ \ }	4.72E-04	\\
13	&{\ \ \ }	4.89E-04	&{\ \ \ }	5.50E-05	&{\ \ \ }	3.45E-04	&{\ \ \ }	5.31E-04	&{\ \ \ }	3.46E-04	\\
14	&{\ \ \ }	3.95E-04	&{\ \ \ }	4.44E-05	&{\ \ \ }	2.60E-04	&{\ \ \ }	4.28E-04	&{\ \ \ }	2.61E-04	\\
15	&{\ \ \ }	3.23E-04	&{\ \ \ }	3.63E-05	&{\ \ \ }	2.02E-04	&{\ \ \ }	3.51E-04	&{\ \ \ }	2.02E-04	\\
16	&{\ \ \ }	2.67E-04	&{\ \ \ }	3.01E-05	&{\ \ \ }	1.60E-04	&{\ \ \ }	2.90E-04	&{\ \ \ }	1.60E-04	\\
`17	&{\ \ \ }	2.24E-04	&{\ \ \ }	2.51E-05	&{\ \ \ }	1.28E-04	&{\ \ \ }	2.43E-04	&{\ \ \ }	1.29E-04	\\
18	&{\ \ \ }	1.89E-04	&{\ \ \ }	2.12E-05	&{\ \ \ }	1.05E-04	&{\ \ \ }	2.05E-04	&{\ \ \ }	1.05E-04	\\
19	&{\ \ \ }	1.61E-04	&{\ \ \ }	1.81E-05	&{\ \ \ }	8.69E-05	&{\ \ \ }	1.75E-04	&{\ \ \ }	8.72E-05	\\
20	&{\ \ \ }	1.38E-04	&{\ \ \ }	1.55E-05	&{\ \ \ }	7.28E-05	&{\ \ \ }	1.50E-04	&{\ \ \ }	7.30E-05	\\
21	&{\ \ \ }	1.20E-04	&{\ \ \ }	1.34E-05	&{\ \ \ }	6.15E-05	&{\ \ \ }	1.30E-04	&{\ \ \ }	6.18E-05	\\
22	&{\ \ \ }	1.04E-04	&{\ \ \ }	1.17E-05	&{\ \ \ }	5.25E-05	&{\ \ \ }	1.13E-04	&{\ \ \ }	5.27E-05	\\
23	&{\ \ \ }	9.11E-05	&{\ \ \ }	1.02E-05	&{\ \ \ }	4.52E-05	&{\ \ \ }	9.89E-05	&{\ \ \ }	4.54E-05	\\
24	&{\ \ \ }	8.02E-05	&{\ \ \ }	9.02E-06	&{\ \ \ }	3.92E-05	&{\ \ \ }	8.71E-05	&{\ \ \ }	3.93E-05	\\
25	&{\ \ \ }	7.10E-05	&{\ \ \ }	7.98E-06	&{\ \ \ }	3.42E-05	&{\ \ \ }	7.71E-05	&{\ \ \ }	3.43E-05	\\
26	&{\ \ \ }	6.32E-05	&{\ \ \ }	7.10E-06	&{\ \ \ }	3.00E-05	&{\ \ \ }	6.86E-05	&{\ \ \ }	3.01E-05	\\
27	&{\ \ \ }	5.64E-05	&{\ \ \ }	6.34E-06	&{\ \ \ }	2.65E-05	&{\ \ \ }	6.13E-05	&{\ \ \ }	2.66E-05	\\
28	&{\ \ \ }	5.06E-05	&{\ \ \ }	5.68E-06	&{\ \ \ }	2.35E-05	&{\ \ \ }	5.49E-05	&{\ \ \ }	2.36E-05	\\
29	&{\ \ \ }	4.55E-05	&{\ \ \ }	5.12E-06	&{\ \ \ }	2.09E-05	&{\ \ \ }	4.94E-05	&{\ \ \ }	2.10E-05	\\
30	&{\ \ \ }	4.11E-05	&{\ \ \ }	4.62E-06	&{\ \ \ }	1.87E-05	&{\ \ \ }	4.47E-05	&{\ \ \ }	1.88E-05	\\
31	&{\ \ \ }	3.73E-05	&{\ \ \ }	4.19E-06	&{\ \ \ }	1.68E-05	&{\ \ \ }	4.05E-05	&{\ \ \ }	1.69E-05	\\
32	&{\ \ \ }	3.39E-05	&{\ \ \ }	3.81E-06	&{\ \ \ }	1.52E-05	&{\ \ \ }	3.68E-05	&{\ \ \ }	1.52E-05	\\
33	&{\ \ \ }	3.09E-05	&{\ \ \ }	3.47E-06	&{\ \ \ }	1.37E-05	&{\ \ \ }	3.36E-05	&{\ \ \ }	1.38E-05	\\
34	&{\ \ \ }	2.82E-05	&{\ \ \ }	3.17E-06	&{\ \ \ }	1.25E-05	&{\ \ \ }	3.07E-05	&{\ \ \ }	1.25E-05	\\
35	&{\ \ \ }	2.59E-05	&{\ \ \ }	2.91E-06	&{\ \ \ }	1.14E-05	&{\ \ \ }	2.81E-05	&{\ \ \ }	1.14E-05	\\
36	&{\ \ \ }	2.38E-05	&{\ \ \ }	2.67E-06	&{\ \ \ }	1.04E-05	&{\ \ \ }	2.58E-05	&{\ \ \ }	1.04E-05	\\
37	&{\ \ \ }	2.19E-05	&{\ \ \ }	2.46E-06	&{\ \ \ }	9.51E-06	&{\ \ \ }	2.38E-05	&{\ \ \ }	9.55E-06	\\
38	&{\ \ \ }	2.02E-05	&{\ \ \ }	2.27E-06	&{\ \ \ }	8.73E-06	&{\ \ \ }	2.20E-05	&{\ \ \ }	8.76E-06	\\
39	&{\ \ \ }	1.87E-05	&{\ \ \ }	2.10E-06	&{\ \ \ }	8.04E-06	&{\ \ \ }	2.03E-05	&{\ \ \ }	8.07E-06	\\
40	&{\ \ \ }	1.73E-05	&{\ \ \ }	1.95E-06	&{\ \ \ }	7.41E-06	&{\ \ \ }	1.88E-05	&{\ \ \ }	7.44E-06	\\
41	&{\ \ \ }	1.61E-05	&{\ \ \ }	1.81E-06	&{\ \ \ }	6.85E-06	&{\ \ \ }	1.75E-05	&{\ \ \ }	6.88E-06	\\
42	&{\ \ \ }	1.50E-05	&{\ \ \ }	1.68E-06	&{\ \ \ }	6.35E-06	&{\ \ \ }	1.63E-05	&{\ \ \ }	6.37E-06	\\
43	&{\ \ \ }	1.39E-05	&{\ \ \ }	1.57E-06	&{\ \ \ }	5.89E-06	&{\ \ \ }	1.52E-05	&{\ \ \ }	5.91E-06	\\
44	&{\ \ \ }	1.30E-05	&{\ \ \ }	1.46E-06	&{\ \ \ }	5.48E-06	&{\ \ \ }	1.41E-05	&{\ \ \ }	5.50E-06	\\
45	&{\ \ \ }	1.22E-05	&{\ \ \ }	1.37E-06	&{\ \ \ }	5.10E-06	&{\ \ \ }	1.32E-05	&{\ \ \ }	5.12E-06	\\
46	&{\ \ \ }	1.14E-05	&{\ \ \ }	1.28E-06	&{\ \ \ }	4.76E-06	&{\ \ \ }	1.24E-05	&{\ \ \ }	4.77E-06	\\
47	&{\ \ \ }	1.07E-05	&{\ \ \ }	1.20E-06	&{\ \ \ }	4.44E-06	&{\ \ \ }	1.16E-05	&{\ \ \ }	4.46E-06	\\
48	&{\ \ \ }	1.00E-05	&{\ \ \ }	1.13E-06	&{\ \ \ }	4.16E-06	&{\ \ \ }	1.09E-05	&{\ \ \ }	4.17E-06	\\
49	&{\ \ \ }	9.42E-06	&{\ \ \ }	1.06E-06	&{\ \ \ }	3.90E-06	&{\ \ \ }	1.02E-05	&{\ \ \ }	3.91E-06	\\
50	&{\ \ \ }	8.86E-06	&{\ \ \ }	9.96E-07	&{\ \ \ }	3.66E-06	&{\ \ \ }	9.63E-06	&{\ \ \ }	3.67E-06	\\
51	&{\ \ \ }	8.35E-06	&{\ \ \ }	9.38E-07	&{\ \ \ }	3.44E-06	&{\ \ \ }	9.07E-06	&{\ \ \ }	3.45E-06	\\
52	&{\ \ \ }	7.88E-06	&{\ \ \ }	8.85E-07	&{\ \ \ }	3.23E-06	&{\ \ \ }	8.56E-06	&{\ \ \ }	3.24E-06	\\
53	&{\ \ \ }	7.44E-06	&{\ \ \ }	8.36E-07	&{\ \ \ }	3.05E-06	&{\ \ \ }	8.08E-06	&{\ \ \ }	3.06E-06	\\
54	&{\ \ \ }	7.03E-06	&{\ \ \ }	7.90E-07	&{\ \ \ }	2.87E-06	&{\ \ \ }	7.64E-06	&{\ \ \ }	2.88E-06	\\
55	&{\ \ \ }	6.66E-06	&{\ \ \ }	7.48E-07	&{\ \ \ }	2.71E-06	&{\ \ \ }	7.23E-06	&{\ \ \ }	2.72E-06	\\
56	&{\ \ \ }	6.30E-06	&{\ \ \ }	7.08E-07	&{\ \ \ }	2.56E-06	&{\ \ \ }	6.85E-06	&{\ \ \ }	2.57E-06	\\
57	&{\ \ \ }	5.98E-06	&{\ \ \ }	6.72E-07	&{\ \ \ }	2.42E-06	&{\ \ \ }	6.49E-06	&{\ \ \ }	2.43E-06	\\
58	&{\ \ \ }	5.67E-06	&{\ \ \ }	6.37E-07	&{\ \ \ }	2.30E-06	&{\ \ \ }	6.16E-06	&{\ \ \ }	2.30E-06	\\
59	&{\ \ \ }	5.39E-06	&{\ \ \ }	6.05E-07	&{\ \ \ }	2.18E-06	&{\ \ \ }	5.85E-06	&{\ \ \ }	2.18E-06	\\
60	&{\ \ \ }	5.12E-06	&{\ \ \ }	5.76E-07	&{\ \ \ }	2.07E-06	&{\ \ \ }	5.57E-06	&{\ \ \ }	2.07E-06	\\
61	&{\ \ \ }	4.87E-06	&{\ \ \ }	5.48E-07	&{\ \ \ }	1.96E-06	&{\ \ \ }	5.30E-06	&{\ \ \ }	1.97E-06	\\
62	&{\ \ \ }	4.64E-06	&{\ \ \ }	5.22E-07	&{\ \ \ }	1.86E-06	&{\ \ \ }	5.04E-06	&{\ \ \ }	1.87E-06	\\
63	&{\ \ \ }	4.42E-06	&{\ \ \ }	4.97E-07	&{\ \ \ }	1.77E-06	&{\ \ \ }	4.81E-06	&{\ \ \ }	1.78E-06	\\
64	&{\ \ \ }	4.22E-06	&{\ \ \ }	4.74E-07	&{\ \ \ }	1.69E-06	&{\ \ \ }	4.58E-06	&{\ \ \ }	1.70E-06	\\
65	&{\ \ \ }	4.03E-06	&{\ \ \ }	4.52E-07	&{\ \ \ }	1.61E-06	&{\ \ \ }	4.37E-06	&{\ \ \ }	1.62E-06	\\
66	&{\ \ \ }	3.85E-06	&{\ \ \ }	4.32E-07	&{\ \ \ }	1.54E-06	&{\ \ \ }	4.18E-06	&{\ \ \ }	1.54E-06	\\
67	&{\ \ \ }	3.68E-06	&{\ \ \ }	4.13E-07	&{\ \ \ }	1.47E-06	&{\ \ \ }	3.99E-06	&{\ \ \ }	1.47E-06	\\
68	&{\ \ \ }	3.52E-06	&{\ \ \ }	3.95E-07	&{\ \ \ }	1.40E-06	&{\ \ \ }	3.82E-06	&{\ \ \ }	1.40E-06	\\
69	&{\ \ \ }	3.37E-06	&{\ \ \ }	3.78E-07	&{\ \ \ }	1.34E-06	&{\ \ \ }	3.66E-06	&{\ \ \ }	1.34E-06	\\
70	&{\ \ \ }	3.22E-06	&{\ \ \ }	3.62E-07	&{\ \ \ }	1.28E-06	&{\ \ \ }	3.50E-06	&{\ \ \ }	1.28E-06	\\
71	&{\ \ \ }	3.09E-06	&{\ \ \ }	3.47E-07	&{\ \ \ }	1.22E-06	&{\ \ \ }	3.35E-06	&{\ \ \ }	1.23E-06	\\
72	&{\ \ \ }	2.96E-06	&{\ \ \ }	3.33E-07	&{\ \ \ }	1.17E-06	&{\ \ \ }	3.22E-06	&{\ \ \ }	1.18E-06	\\
73	&{\ \ \ }	2.84E-06	&{\ \ \ }	3.19E-07	&{\ \ \ }	1.12E-06	&{\ \ \ }	3.09E-06	&{\ \ \ }	1.13E-06	\\
74	&{\ \ \ }	2.73E-06	&{\ \ \ }	3.06E-07	&{\ \ \ }	1.08E-06	&{\ \ \ }	2.96E-06	&{\ \ \ }	1.08E-06	\\
75	&{\ \ \ }	2.62E-06	&{\ \ \ }	2.94E-07	&{\ \ \ }	1.03E-06	&{\ \ \ }	2.85E-06	&{\ \ \ }	1.04E-06	\\
76	&{\ \ \ }	2.52E-06	&{\ \ \ }	2.83E-07	&{\ \ \ }	9.91E-07	&{\ \ \ }	2.73E-06	&{\ \ \ }	9.95E-07	\\
77	&{\ \ \ }	2.42E-06	&{\ \ \ }	2.72E-07	&{\ \ \ }	9.52E-07	&{\ \ \ }	2.63E-06	&{\ \ \ }	9.56E-07	\\
78	&{\ \ \ }	2.33E-06	&{\ \ \ }	2.61E-07	&{\ \ \ }	9.15E-07	&{\ \ \ }	2.53E-06	&{\ \ \ }	9.18E-07	\\
79	&{\ \ \ }	2.24E-06	&{\ \ \ }	2.52E-07	&{\ \ \ }	8.80E-07	&{\ \ \ }	2.43E-06	&{\ \ \ }	8.83E-07	\\
80	&{\ \ \ }	2.16E-06	&{\ \ \ }	2.42E-07	&{\ \ \ }	8.46E-07	&{\ \ \ }	2.34E-06	&{\ \ \ }	8.49E-07	\\
\end{longtable}

In view of the above comparisons of various works with the results of Amin et al.\cite{Amin2008}, and their poor agreement with theory, serious concerns can be raised about the overall accuracy of the data produced by the experimental methods described in that paper.
It is noteworthy that with a revised, and accurate, measurement on the $7s$ state, the work of Yar et al.\cite{Yar2013} represents a step in remedying that situation. The work of Kalyar et al.\cite{Kalyar2016} improves the discussion on the $4p$ photoionization but indicates a large difference between the $4p_{1/2}$ and $4p_{3/2}$ states, much bigger than the theoretical approaches indicate. Furthermore, the latter work's results on absorption oscillator strengths are not overall ``well behaved,'' and need to be reevaluated.

\end{document}